\documentclass[conference]{IEEEtran}
\IEEEoverridecommandlockouts
\usepackage{graphicx}
\usepackage{amsmath}
\usepackage{balance}
\usepackage{multirow}
\usepackage{cite}
\usepackage{stmaryrd}
\usepackage[hidelinks]{hyperref}
\usepackage{balance}

\def\BibTeX{{\rm B\kern-.05em{\sc i\kern-.025em b}\kern-.08em
    T\kern-.1667em\lower.7ex\hbox{E}\kern-.125emX}}
\begin{document}

\title{Neural network-based acoustic vehicle counting\\
\thanks{The research was supported by Research Center for Informatics (project CZ.02.1.01/0.0/0.0/16019/0000765 funded by OP VVV) and CTU student grant (SGS OHK3-019/20). Slobodan Djukanovi\'{c} was supported by the OP RDE programme of project International Mobility of Researchers MSCA-IF III at CTU in Prague No. CZ.02.2.69/0.0/0.0/19\_074/0016255.}
}

\author{\IEEEauthorblockN{1\textsuperscript{st} Slobodan Djukanovi\'{c}}
\IEEEauthorblockA{\textit{Faculty of Electrical Engineering} \\
\textit{University of Montenegro}\\
Podgorica, Montenegro \\
slobdj@ucg.ac.me}
\and
\IEEEauthorblockN{2\textsuperscript{nd} Yash Patel}
\IEEEauthorblockA{\textit{Faculty of Electrical Engineering} \\
\textit{Czech Technical University}\\
Prague, Czech Republic \\
patelyas@fel.cvut.cz}
\and
\IEEEauthorblockN{3\textsuperscript{rd} Ji\v{r}\'{i} Matas}
\IEEEauthorblockA{\textit{Faculty of Electrical Engineering} \\
\textit{Czech Technical University}\\
Prague, Czech Republic \\
matas@fel.cvut.cz}
\and
\IEEEauthorblockN{4\textsuperscript{th} Tuomas Virtanen}
\IEEEauthorblockA{\textit{Faculty of Information Technology and Communication Sciences} \\
\textit{Tampere University}\\
Tampere, Finland \\
tuomas.virtanen@tuni.fi}
}
\maketitle

\begin{abstract}
This paper addresses acoustic vehicle counting using one-channel audio. We predict the pass-by instants of vehicles from local minima of clipped vehicle-to-microphone distance. This distance is predicted from audio using a two-stage (coarse-fine) regression, with both stages realised via neural networks (NNs). Experiments show that the NN-based distance regression outperforms by far the previously proposed support vector regression.
The $ 95\% $ confidence interval for the mean of vehicle counting error is within $[0.28\%, -0.55\%]$.
Besides the minima-based counting, we propose a deep learning counting that operates on the predicted distance without detecting local minima.
Although outperformed in accuracy by the former approach, deep counting has a significant advantage in that it does not depend on minima detection parameters.
Results also show that removing low frequencies in features improves the counting performance.
\end{abstract}


\section{Introduction}\label{Introduction}

Traffic monitoring (TM) systems use different traffic data to improve the use and performance of roadway systems, transportation safety, law enforcement, prediction of future transportation needs etc. TM data include estimates of vehicle count, traffic volume, speed of vehicles and of various vehicle parameters (length, weight, class) \cite{won2020intelligent}.

Current TM systems use diverse sensors and technologies, including induction loops, vibration, piezoelectric, infrared, ultrasonic, magnetic and acoustic sensors and cameras \cite{won2020intelligent}.
Vision-based TM systems have recently become popular due to breakthroughs in object detection, tracking and classification tasks provided by deep learning methods \cite{naphade2019}. In addition, a single camera suffices to cover multiple lanes for the TM tasks, which is not the case for other sensor technologies.
However, high computational complexity, partial occlusion, shadows and illumination variation limit the performance of vision-based TM systems \cite{won2020intelligent,morris2008survey}.

Acoustic TM has several advantages in comparison with the vision-based one \cite{won2020intelligent}. Microphones are less expensive than cameras, consume less energy, require less storage space, are easier to install and maintain with low wear and tear. In addition, acoustic TM is not affected by visual occlusions and lighting conditions, and has less privacy issues.

This paper addresses acoustic vehicle counting using one-channel audio. The standard approach is to detect temporal variation of the sound power due to vehicles passing by microphone \cite{kato2005attempt,george2013vehicle,george2013exploring}, which is performed using state transitions of a hidden Markov model \cite{kato2005attempt} or by a peak-picking algorithm \cite{george2013vehicle,george2013exploring}. Maximal frequency at which the power of a time-frequency representation reaches a predefined threshold, a.k.a. top-right frequency, also enables detecting vehicles passing by the microphone \cite{li2017auto++}. The method \cite{djukanovic2020robust} uses a prediction of \textit{clipped vehicle-to-microphone distance}, a form of pseudo-distance between a vehicle and the microphone.
Vehicle counting \cite{djukanovic2020robust}, carried out by counting local minima in the predicted distance, outperforms those based on peak detection in the sound power and top-right frequency.
However, the optimal, false negative-false positive compensating, detection threshold in \cite{djukanovic2020robust} can only be imprecisely estimated \textit{a priori}. Moreover, the distance regression \cite{djukanovic2020robust} is computationally demanding.

In this paper, we significantly improve the distance regression, and thus vehicle counting accuracy, compared with \cite{djukanovic2020robust} by a computationally less demanding approach. We first overview clipped vehicle-to-microphone distance (Section \ref{CVMD_section}) and then propose new counting method (Section \ref{ProposedMethod}). Experimental results are given in Section \ref{Experiment}, whereas Section \ref{Conclusions} concludes the paper.

\section{Clipped vehicle-to-microphone distance and vehicle counting}\label{CVMD_section}

\begin{figure}[th!]
	\centering
	\includegraphics{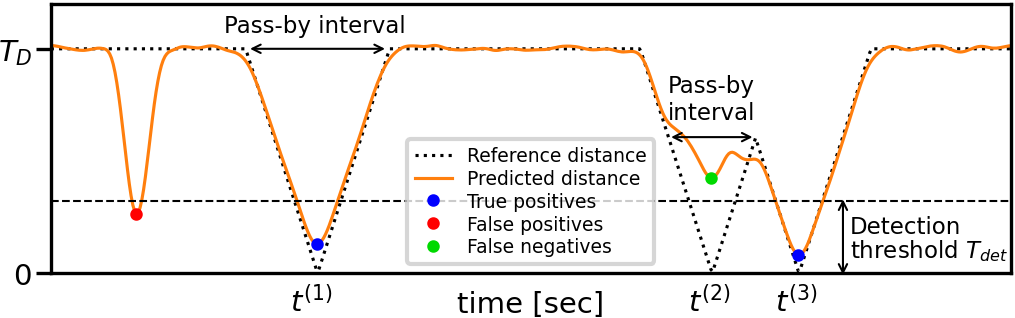}
	\caption{Illustration of a reference distance, distance predicted from audio and classification of predicted distance minima.}
	\label{Fig1}
\end{figure}

In \cite{djukanovic2020robust}, clipped vehicle-to-microphone distance\footnote{We will refer to clipped vehicle-to-microphone distance as the distance.} of the $ k$-th vehicle is defined as
\begin{equation}\label{distance}
	d^{(k)}(t)=\begin{cases}
		\left|t-t^{(k)}\right|,	& \quad \left|t-t^{(k)}\right| < T_D\\
		T_D, & \quad  \text{elsewhere},
	\end{cases}
\end{equation}
where $ t^{(k)} $ represents the pass-by instant of the vehicle and $ T_D $ is the distance threshold. The V-shape of $d^{(k)}(t)$ models approaching and receding of the vehicle from the microphone (see dotted line around $t^{(1)}$ in Fig. \ref{Fig1}). When the audio contains $ N_{v} $ vehicles, only the distance of the closest vehicle is taken into account, so the overall distance is defined as minimum of all separate distances (dotted line in Fig. \ref{Fig1}):
\begin{equation}\label{CVMD}
	D(t)=\min\{d^{(1)}(t),d^{(2)}(t),\dots, d^{(N_{v})}(t)\}.
\end{equation}

In \cite{djukanovic2020robust}, the vehicle count is equal to the number of detected local minima of the predicted distance (orange line in Fig. \ref{Fig1}) that fall below detection threshold. Not every local minimum below the threshold corresponds to a vehicle passing by the microphone. Only minima that occur within the true vehicle pass-by intervals (depicted by horizontal arrows in Fig. \ref{Fig1}) represent true positives (TPs). Other minima below the threshold represent false positives (FPs), whereas minima that occur within the corresponding pass-by intervals, but are above the threshold, represent false negatives (FNs).

The method \cite{djukanovic2020robust} is evaluated using the TP, FP and FN probabilities, $ p_{\text{TP}} $, $ p_{\text{FP}} $ and $ p_{\text{FN}} $, calculated for variable detection threshold. The optimal threshold corresponds to a point where $ p_{\text{FP}} = p_{\text{FN}} $, since then the FPs and FNs cancel each other in statistical sense and the total number of detected vehicles equals the true number of vehicles. In terms of counting error, the best generalization in \cite{djukanovic2020robust} is obtained when the distance is predicted using the log-mel spectrogram (LMS) and newly introduced high-frequency power (HFP) as input features. With HFP, the counting error remains low (below $ 2\% $) within a wide range of detection threshold values.

Although characterized by low counting error within a wide range of detection thresholds, the optimal threshold of \cite{djukanovic2020robust} is not known in advance. Our first objective is to extend threshold range with low-error (below $ 2\% $ or even more), i.e., to make counting more robust to the choice of detection threshold. Another drawback of \cite{djukanovic2020robust} is the computational complexity. For distance regression, it uses support vector regression (SVR), implemented in the libsvm library \cite{chang2011libsvm}. Its complexity scales between $ O(n_{f} n^2_{s}) $ and $ O(n_{f} n^3_{s}) $, where $ n_{f} $ and $ n_{s} $ represent the number of features and samples in the dataset, respectively. Therefore, our second objective is to perform distance regression in a computationally more efficient way to enable scaling to larger datasets. A method which fulfills these two objectives is described in the sequel.

\section{Neural network-based counting}\label{ProposedMethod}

\begin{figure}[th!]
	\centering
	\includegraphics{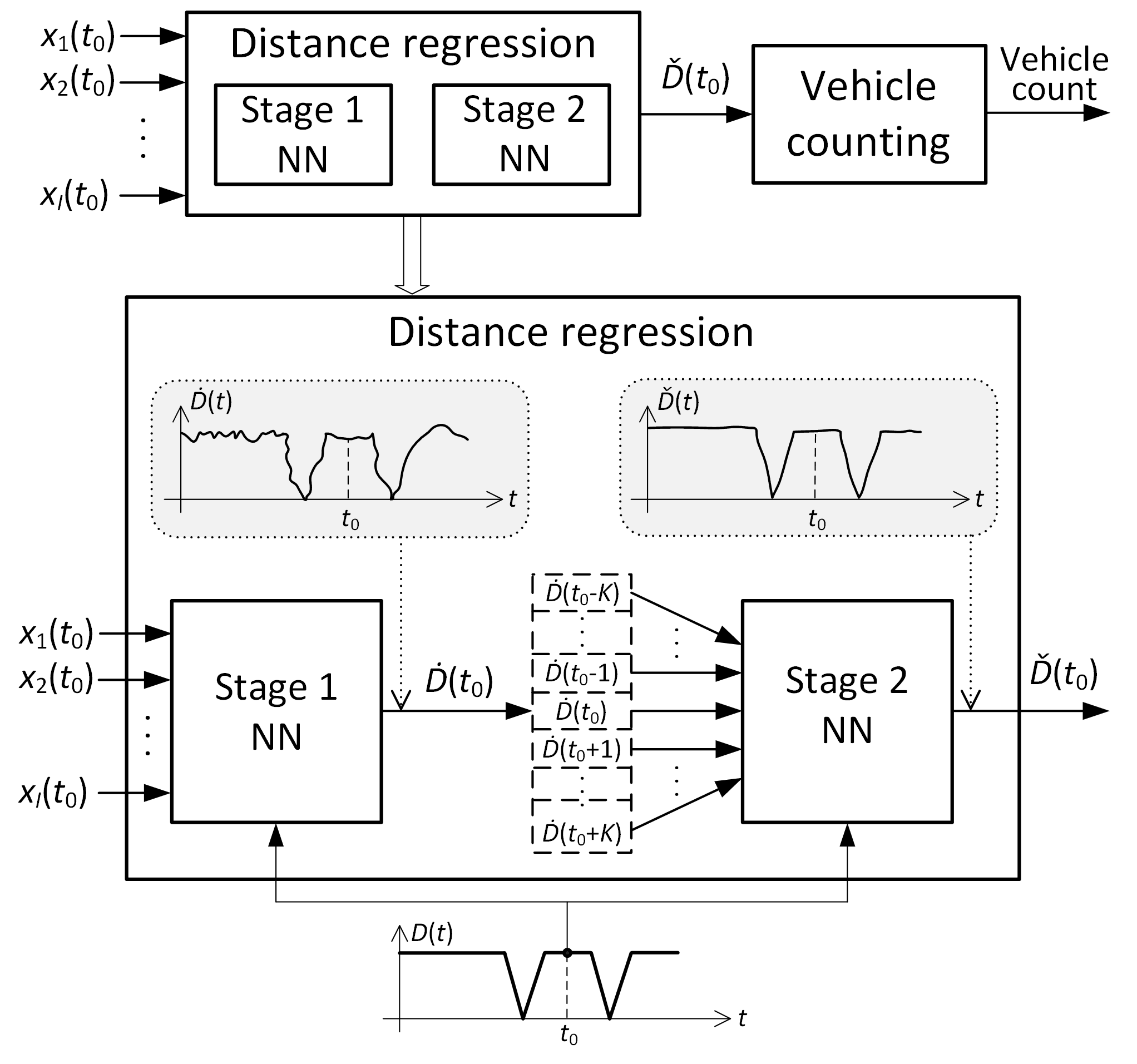}
	\caption{\textit{Top}: The block diagram of the proposed vehicle counting method. \textit{Bottom}: Distance regression in detail. Stage 2 improves the distance regression output by Stage 1.}
	\label{Fig2}
\end{figure}

To address the low-counting error objective, we propose to improve the accuracy of distance regression by a two-stage (coarse-fine) approach. To address the computational complexity objective, we propose to use fully-connected neural networks (NNs) instead of originally used SVR. The block diagram of the proposed method is presented in Fig. \ref{Fig2} (top). $ x_i(t_0) $, $ i=1,\cdots,I $ are input features at time instant $ t_0 $ and $\check{D}(t_{0}) $ represents the predicted distance at $ t_0 $. The \textit{Vehicle counting} block carries out counting based on the distance prediction of the whole audio file.

\subsection{Distance regression and input features}\label{CVMD_regression}

A detailed representation of the proposed distance regression is given in Fig. \ref{Fig2} (bottom). Stage 1 NN performs regression based on input features $ x_i(t_0) $, $ i=1,\cdots,I $, similarly to SVR in \cite{djukanovic2020robust}. As input features, we propose to combine HFP and LMS, as suggested in \cite{djukanovic2020robust}. Since HFP represents the power of high-frequency portion of the signal spectrum, we incorporate it into LMS by leaving out a number of filters with the lowest central frequencies in the mel spectrogram filter bank \cite{serizel2018acoustic}. The resulting LMS, referred to as high-frequency LMS (HF-LMS), does not include low-frequency portion of the spectrum which contains the most significant part of the environment noise. To take into account time dependence between adjacent $ D(t) $ values, value $ D(t_0) $ will be predicted using the samples of the HF-LMS spectrum from time interval $ [t_0-Q,t_0+Q ]$ . Therefore, the dimensionality of the input space is $I=(2Q+1)N_{mel}$, where $N_{mel}$ is the number of mel bands. For implementation details, see Section \ref{ImpDetails}.

The task of Stage 2 NN is to refine prediction $ \dot{D}(t_0) $ of Stage 1 to obtain the final distance prediction $ \check{D}(t_0) $. Stage 2 refines $ \dot{D}(t_0) $ using a window of predicted distances centered at $ t_0 $, i.e., a vector of $ 2K+1 $ successive predicted distances $ \dot{D}(t_0-K), \dots, \dot{D}(t_0+K)$, represents input features to Stage 2 NN.

Prior to vehicle counting, the predicted distance is low-pass filtered to eliminate high-frequency oscillations, which is discussed in Section \ref{Experiment}.

\subsection{Vehicle counting}\label{CVMD_min_det}
In this paper, we propose two vehicle counting approaches, both based on the final (Stage 2) predicted distance. 

\subsubsection{Peak properties-based vehicle counting}\label{CVMD_min_det_PP}
\begin{figure}[th!]
	\centering
	\includegraphics{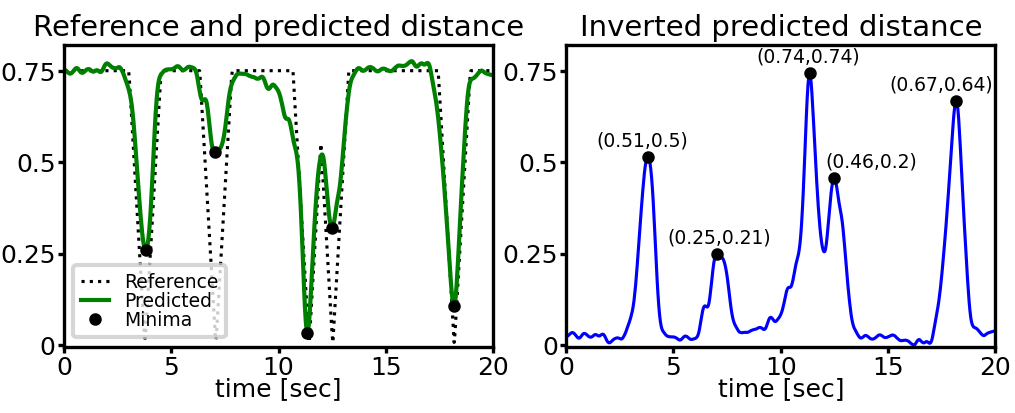}
	\caption{\textit{Left}: Reference and predicted distances of an audio file. \textit{Right}: Inverted predicted distance is used for vehicle detection. Each detected peak (black dots) is characterized by a magnitude-prominence pair (e.g., first peak $ (0.51,0.5) $).}
	\label{Fig3}
\end{figure}

In \cite{djukanovic2020robust}, local minima of the predicted distance (black dots in Fig. \ref{Fig3} (left)) were detected by detecting local maxima (peaks) of the inverted distance (black dots in Fig. \ref{Fig3} (right)) based on their prominence\footnote{The prominence of a peak measures how much the peak stands out due to its height and location relative to other peaks, and is defined as the vertical distance between the peak and its lowest contour line \cite{peakprominence}.}. Here, we extend this approach by introducing a peak magnitude criterion. If two close peaks have similar magnitudes (corresponding to two close vehicles), the prominence of the weaker one can be much less than that of the stronger one. For example, consider the fourth peak with  $0.46$ magnitude and $0.2$ prominence, shortly $(0.46,0.2) $, in Fig. \ref{Fig3} (right). It clearly corresponds to a vehicle (see the corresponding reference and predicted distances in Fig. \ref{Fig3} (left)), but due to vicinity of the third peak $ (0.74,0.74) $ its prominence is even smaller than that of the second peak $ (0.25,0.21) $ which has almost two times smaller magnitude.

Since the weaker peak can be left out if it is detected based on prominence only, we define vehicle detection criterion as: 

\vspace{1mm}
\leftskip0.6cm\relax
\rightskip0.6cm\relax
\noindent\textit{A vehicle is detected if detected peak of the inverted predicted distance has magnitude larger than $ M $ or prominence larger than $ P $.}

\leftskip0cm\relax
\rightskip0cm\relax
\vspace{1mm}

\noindent Selection of $ M $ and $ P $ is discussed in Section \ref{Experiment}.

\subsubsection{Deep vehicle counting}
\label{deep_peak_det}

A convolutional NN \cite{zhang2015character} is used for counting. The model operates on the raw predicted distance and estimates the vehicle count directly, without an intermediate local minima detection. It consists of 1D convolutional, a global-average pooling and fully-connected layers. The global-average-pooling allows the model to operate on varying length distance \cite{lin2013network}. 

The model is trained to predict the vehicle count directly. We experimented with three different loss functions: $L_{1}$, $L_{2}$ and smooth $L_{0}$ distances. The model trained with smooth $L_{0}$ distance gives the best performance. Training with smooth $L_{0}$ is performed using a surrogate learned via a deep embedding, where the Euclidean distance between the prediction and the ground truth corresponds to the $L_{0}$ distance \cite{patel2020learning}. The deep embedding is realized using a shallow fully-connected NN. The surrogate and the counting model are trained in parallel.

\subsection{Implementation details}
\label{ImpDetails}
The HF-LMS feature is based on the spectrogram of the input signal. In this paper, the length of sliding (Hamming) window used in spectrogram calculation is $ N_w=4096 $ and the stride length is $ N_h=1634 $ samples \cite{stankovic2014instantaneous}, which with $20$-second audio files sampled at $ f_s=44100 $ Hz\footnote{The spectrogram of audio file presented in Fig. 3 in \cite{djukanovic2020robust} indicates the presence of high-frequency components (around $ 14 $ kHz). Therefore, the original sampling rate is retained.} gives the time-length of HF-LMS of $ 540 $ samples. In addition, $ N_{mel}=48 $ mel bands are used in HF-LMS, with $f_{min}=1000$ Hz and $ f_{max} = fs/2 =22050$ Hz. To form a vector of input features, we take the HF-LMS spectra at $ Q=5 $ preceding and following instants with a stride of $ 2 $. Therefore, the dimensionality of the input space is $I=(2Q+1)N_{mel}=528$. The input dimensionality of Stage 2 NN is $ 2K+1=31$. For distance threshold, we take $ T_D=0.75 $ s \cite{djukanovic2020robust}. The selected values of $ N_{mel} $, $ Q $, $ K $ and $ T_D $ are obtained via cross-validation. 

Stage 1 and 2 NNs have four layers with $ 528$-$ 64 $-$ 64 $-$1 $ and $ 31$-$ 31 $-$ 15 $-$1 $ neurons per layer, respectively. These configurations gave the best regression performance cross-validated on the training data. Both NNs use mean squared error loss, ReLU activation, $ L2 $ kernel regularization with factors $ 10^{-4} $ (Stage 1) and $ 5\times 10^{-6} $ (Stage 2), $ 100 $ training epochs. Batch normalization is applied at each layer, except for the output, after activation. The model is implemented in Keras. Peak detection is based on \texttt{scipy.signal.find\_peaks} procedure (SciPy library for Python).

\section{Experiments}\label{Experiment}

First evaluation metric we use is {\it{normalized area under the}} $ p_{\text{TP}} $ {\it{curve}}, i.e., the average value of $ p_{\text{TP}}$ over detection threshold $T_{det}\in[0,T_{D}]$ \cite{djukanovic2020robust}. $ p_{\text{TP}}$ is calculated as percentage of minima detected within the pass-by intervals (maximum one minimum per interval) at $100$ equidistant $T_{det}$ points.

The second metric is \textit{relative vehicle counting error}
\begin{equation}\label{RVCE}
	\text{RVCE}=\frac{N_{v}^{true}-N_{v}^{est}}{N_{v}^{true}} \times 100 \,[\%],
\end{equation}
where $ N_{v}^{true} $ and $ N_{v}^{est} $ represent the true and estimated vehicle counts. As opposed to (\ref{RVCE}), the error definition in \cite{djukanovic2020robust} uses the absolute difference. Here, the signed error enables distinguishing between underestimation (positive error) and overestimation (negative error) in counting.

We will use the dataset from \cite{djukanovic2020robust}\footnote{The dataset and Python implementation of \cite{djukanovic2020robust} are available for download at \url{http://cmp.felk.cvut.cz/data/audio_vc}.} which contains two parts: VC-PRG-1:5 ($250$ $20$-second audio files with $841$ vehicles) and VC-PRG-6 ($172$ audio files with $580$ vehicles). Five vehicle classes are considered in the dataset: motorcycles, cars, vans, buses and trucks. No assumptions are made regarding the vehicles' speed, nor it is estimated.

The proposed vehicle counting method (hereafter referred to as VCNN) will be trained and validated ($80\%$-$20\%$ training-validation split) using VC-PRG-1:5\footnote{With the setup given in Section \ref{ImpDetails}, the number of features and samples in the training/validation dataset are $ n_f =528$ and $ n_s=250\times 540=135000 $, respectively. Since the SVR complexity scales between $ O(n_{f} n^2_{s}) $ and $ O(n_{f} n^3_{s}) $ \cite{chang2011libsvm}, we have decided to implement NN-based regression.} and tested on VC-PRG-6. We run the method $ 40 $ times (training data shuffled each time). Along with Stage 2, we also consider the output of Stage 1 NN and the corresponding vehicle count VCNN$_{S1}$.

Minima detection, and therefore probabilities $ p_{\text{TP}} $, $ p_{\text{FP}} $ and $ p_{\text{FN}} $, as well as RVCE, are affected by i) low-pass filtering of the predicted distance, ii) value of $ M$ and iii) value of $ P$ (Section \ref{CVMD_min_det_PP}). To optimize the count performance with respect to these factors, we carry out the grid search algorithm. We consider all possible combinations of i) low-pass filters (successive moving average (MA) filters with lengths $ (5,3) $\footnote{($l_1,l_2$) denotes filtering first with an MA filter of length $l_1$ then with one of length $l_2$.}, $ (7,3) $ and $ (7,5,3) $), ii) $ M\in\{35\%\,T_D,40\%\,T_D,45\%\,T_D,50\%\,T_D\} $ and iii) $ P\in\{10\%\,T_D,15\%\,T_D,20\%\,T_D,25\%\,T_D\} $. The optimality criterion is averaged absolute RVCE for detection threshold range $ T_{det}\in[50\%\,T_D, T_D] $. Table \ref{Tab1} presents the optimal setups for VCNN$_{S1}$ and VCNN (first two rows).
\vspace{-1mm}

\begin{table}[h]
	\centering
	\caption{Optimal minima (vehicle) detection parameters}
	\label{Tab1}
	\begin{tabular}{ l | c }
		\hline
		\multicolumn{1}{c|}{Setup} & \multicolumn{1}{c}{Optimal parameters} \\ \hline
		VCNN$_{S1}$ & MA filters$ \, (7,3)$, $M= 45\%\,T_D$, $P= 25\%\,T_D$\\
		VCNN & MA filters$ \, (5,3)$, $M= 40\%\,T_D$, $P= 20\%\,T_D$\\
		VCNN$_{f0} $ & MA filters$ \, (5,3)$, $M= 45\%\,T_D$, $P= 20\%\,T_D$\\
		\hline
	\end{tabular}
\end{table}

\begin{figure}[th!]
	\centering
	\includegraphics{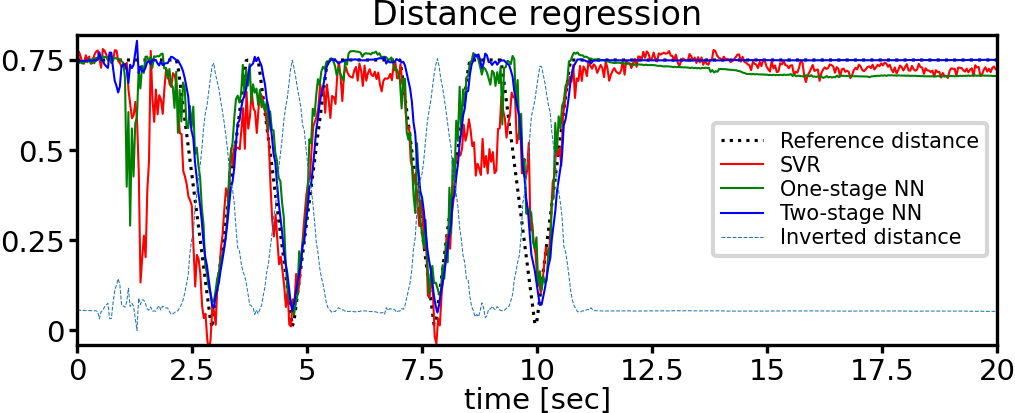}
	\caption{Distance predictions of an audio file. Minima are detected by detecting peaks of the inverted predicted distance.}
	\label{Fig4}
\end{figure}

\begin{figure}[th!]
	\centering
	\includegraphics{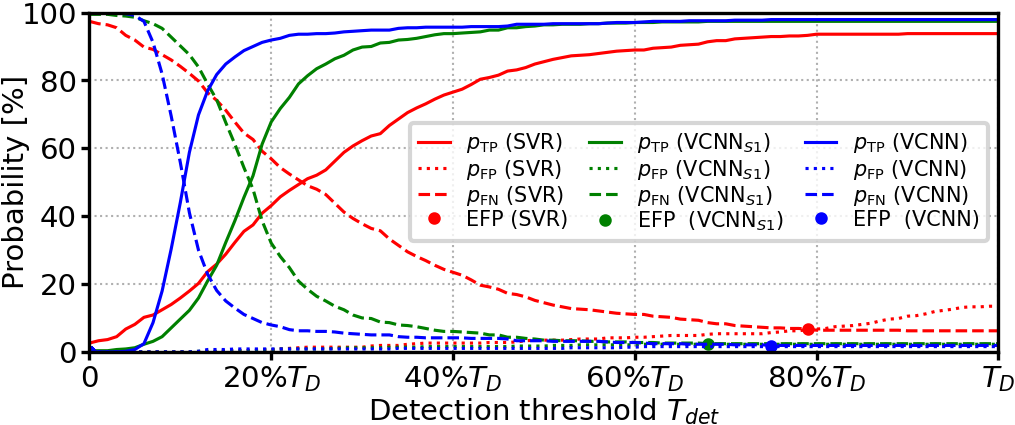}
	\caption{TP, FP and FN probabilities with equal false probabilities (EFP) points. The lower EFP the better.}
	\label{Fig5}
\end{figure}

Before evaluating the vehicle counting performance, let's analyze improvement in distance regression offered by the proposed approach. Figure \ref{Fig4} compares distance predictions of one audio file carried out via SVR, one-stage NN and two-stage NN. The two-stage approach (blue solid line) follows the reference distance (dotted line) the most faithfully. The mean square regression errors for the testing set are $ 9.55 \times 10^{-3}$ (SVR), $ 5.27 \times 10^{-3}$ (one-stage NN) and $ 2.30 \times 10^{-3}$ (two-stage NN). We conclude that i) the proposed method significantly outperforms SVR in terms of regression accuracy, and ii) Stage 2 improves the regression accuracy with respect to Stage 1.

Figure \ref{Fig5} compares probabilities $ p_{\text{TP}} $, $ p_{\text{FP}} $ and $ p_{\text{FN}} $ of the SVR-based counting \cite{djukanovic2020robust} and of one run of VCNN$_{S1}$ and VCNN. Low $ T_{det} $ results in low $ p_{\text{TP}} $ and high $ p_{\text{FN}} $ (see detection threshold in Fig. \ref{Fig1}). On the other hand, with high $ T_{det} $, majority of detected local minima surpass the threshold, resulting in high $ p_{\text{TP}} $ and low $ p_{\text{FN}} $. Here, we notice notably higher $ p_{\text{FP}} $ (dotted curves in Fig. \ref{Fig5}) in the SVR-based counting. Significant improvement of the proposed approach is reflected in an increase of normalized area under the $ p_{\text{TP}} $ curve from $ 0.695 $ (SVR) to $ 0.759 $ (VCNN$_{S1}$) and $ 0.850 $ (VCNN), the latter two averaged over all runs. Dots in Fig. \ref{Fig5} represent the points of equal false probabilities (EFPs), which are $ 6.55\% $, $ 2.33\% $ and $ 1.72\% $, respectively. The lower EFP the better \cite{djukanovic2020robust}.

\begin{figure}[t!]
	\centering
	\includegraphics{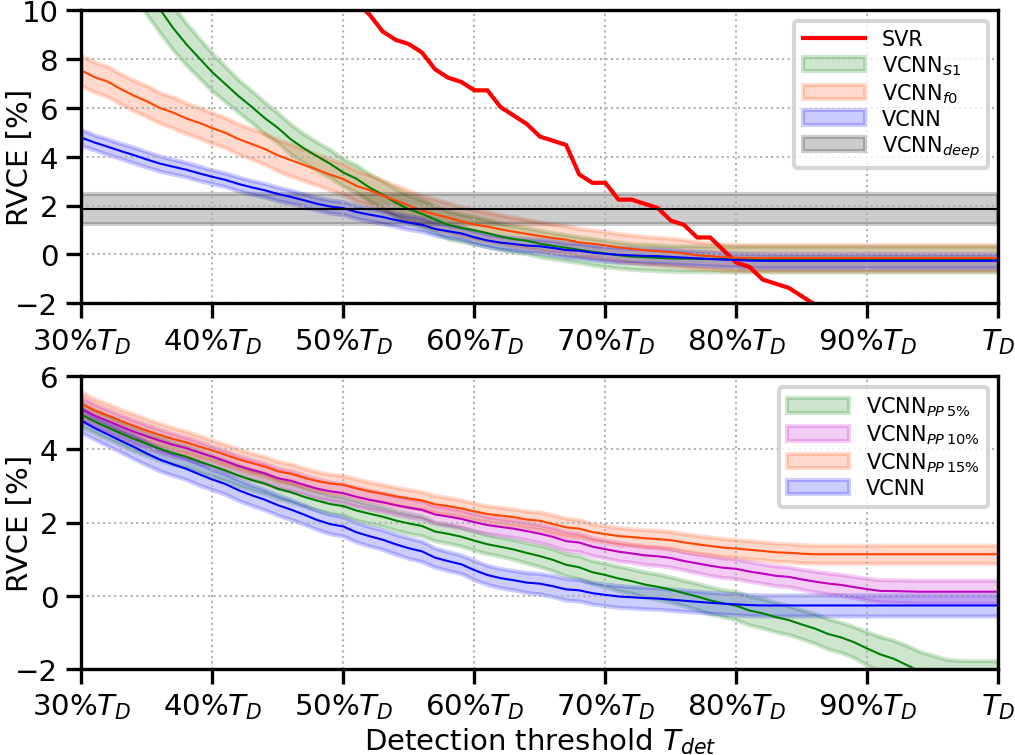}
	\caption{Relative vehicle counting error, RVCE, as a function of the detection threshold. With the exception of SVR, RVCEs are shown as $95\%$ confidence intervals for the mean. The proposed VCNN (in blue) compared with alternatives (for their description, see text). VCNN$_{PP\,\#\%}$ in the bottom plot correspond to peak detection based only on peak prominence.}
	\label{Fig6}
\end{figure}

The error plots for detection threshold $T_{det}\in [30\%\,T_D,T_D] $ are given in Fig. \ref{Fig6} (top). With the exception of SVR, RVCEs are shown as $ 95\% $ confidence intervals for the mean (referred to as confidence). In addition to vehicle counting based on SVR, VCNN$_{S1}$ and VCNN, we present RVCE of the deep counting approach (VCNN$_{deep}$) and of VCNN when full-frequency range is used in the LMS input features ($ f_{min}=0 $, see Section \ref{ImpDetails}). The latter is denoted as VCNN$_{f0}$ and its optimal parameters are given in the bottom row in Table \ref{Tab1}. Confidence of VCNN$_{deep}$ is a horizontal band within $[1.25\%, 2.49\%]$. The deep counting does not depend on minima detection parameters and hence its performance does not depend on $T_{det}$. 
The proposed VCNN approach outperforms the other ones by a significant margin. Its $ 95\% $ confidence goes below $ 2\% $ for $T_{det}\approx 50\%\,T_D $ and is bounded within $[0.28\%, -0.55\%]$ for $T_{det} > 70\%\,T_D $. In other words, the counting error is less then $ 2\% $ for $T_{det} > 50\%\,T_D $, which makes our low-error objective fulfilled.

The combined peak magnitude-prominence criterion in vehicle detection (Section \ref{CVMD_min_det_PP}) improves the counting performance with respect to detection based solely on the peak prominence \cite{djukanovic2020robust}, as illustrated in Fig. \ref{Fig6} (bottom). Peak prominences of $ 5\% $, $ 10\% $ and $ 15\% $ are considered. Prior to peak detection, the predicted distance has been low-pass filtered with MA filters $(5,3)$. Introducing magnitude in peak detection is crucial for achieving stable accurate results within a wide range of detection threshold. 

\section{CONCLUSIONS}\label{Conclusions}
We proposed a method for acoustic vehicle counting based on the clipped vehicle-to-microphone distance. The distance was predicted using a two-stage NN-based regression. Significant improvement in regression accuracy with respect to the SVR-based approach resulted in a highly accurate vehicle counting not depending on detection threshold within a wide range of threshold values. Deep counting, an alternative to the local minima-based counting, estimates the vehicle count directly from the predicted distance, without detecting local minima. Although outperformed in accuracy by the latter approach, a significant advantage of deep counting is that it does not depend on minima detection parameters. Our future work will address developing end-to-end vehicle counting method.

\balance

\bibliographystyle{IEEEbib}
\bibliography{References}

\begin{thebibliography}{10}

\bibitem{won2020intelligent}
Myounggyu Won,
\newblock ``Intelligent traffic monitoring systems for vehicle classification:
  A survey,''
\newblock {\em IEEE Access}, vol. 8, pp. 73340--73358, 2020.

\bibitem{naphade2019}
Milind~Naphade et~al.,
\newblock ``The 2019 {AI} {C}ity challenge,''
\newblock in {\em CVPR Workshops}, 2019, pp. 452--460.

\bibitem{morris2008survey}
Brendan~Tran Morris and Mohan~Manubhai Trivedi,
\newblock ``A survey of vision-based trajectory learning and analysis for
  surveillance,''
\newblock {\em IEEE Transactions on Circuits and Systems for Video Technology},
  vol. 18, no. 8, pp. 1114--1127, 2008.

\bibitem{kato2005attempt}
Jien Kato,
\newblock ``An attempt to acquire traffic density by using road traffic
  sound,''
\newblock in {\em Proceedings of the 2005 International Conference on Active
  Media Technology, 2005.(AMT 2005).} IEEE, 2005, pp. 353--358.

\bibitem{george2013vehicle}
Jobin George, Leena Mary, and KS~Riyas,
\newblock ``Vehicle detection and classification from acoustic signal using
  {ANN} and {KNN},''
\newblock in {\em 2013 international conference on control communication and
  computing (ICCC)}. IEEE, 2013, pp. 436--439.

\bibitem{george2013exploring}
Jobin George, Anila Cyril, Bino~I Koshy, and Leena Mary,
\newblock ``Exploring sound signature for vehicle detection and classification
  using {ANN},''
\newblock {\em International Journal on Soft Computing}, vol. 4, no. 2, pp. 29,
  2013.

\bibitem{li2017auto++}
Sugang Li, Xiaoran Fan, Yanyong Zhang, Wade Trappe, Janne Lindqvist, and
  Richard~E Howard,
\newblock ``Auto++: Detecting cars using embedded microphones in real-time,''
\newblock {\em Proceedings of the ACM on Interactive, Mobile, Wearable and
  Ubiquitous Technologies}, vol. 1, no. 3, pp. 70, 2017.

\bibitem{djukanovic2020robust}
Slobodan Djukanovi{\'c}, Ji{\v{r}}{\'\i} Matas, and Tuomas Virtanen,
\newblock ``Robust audio-based vehicle counting in low-to-moderate traffic
  flow,''
\newblock in {\em 2020 IEEE Intelligent Vehicles Symposium (IV)}. IEEE, 2020,
  pp. 1608--1614.

\bibitem{chang2011libsvm}
Chih-Chung Chang and Chih-Jen Lin,
\newblock ``{LIBSVM}: A library for support vector machines,''
\newblock {\em ACM Transactions on Intelligent Systems and Technology (TIST)},
  vol. 2, no. 3, pp. 1--27, 2011.

\bibitem{serizel2018acoustic}
Romain Serizel, Victor Bisot, Slim Essid, and Ga{\"e}l Richard,
\newblock ``Acoustic features for environmental sound analysis,''
\newblock in {\em Computational Analysis of Sound Scenes and Events}, pp.
  71--101. Springer, 2018.

\bibitem{peakprominence}
``Topographic prominence,''
  \url{https://en.wikipedia.org/wiki/Topographic_prominence},
\newblock Accessed: 2020-10-19.

\bibitem{zhang2015character}
Xiang Zhang, Junbo Zhao, and Yann LeCun,
\newblock ``Character-level convolutional networks for text classification,''
\newblock in {\em NeurIPS}, 2015.

\bibitem{lin2013network}
Min Lin, Qiang Chen, and Shuicheng Yan,
\newblock ``Network in network,''
\newblock {\em ICLR}, 2014.

\bibitem{patel2020learning}
Yash Patel, Tom{\'a}{\v{s}} Hoda{\v{n}}, and Ji{\v{r}}{\'\i} Matas,
\newblock ``Learning surrogates via deep embedding,''
\newblock in {\em European Conference on Computer Vision}. Springer, 2020, pp.
  205--221.

\bibitem{stankovic2014instantaneous}
Ljubi{\v{s}}a Stankovi{\'c}, Igor Djurovi{\'c}, Srdjan Stankovi{\'c}, Marko
  Simeunovi{\'c}, Slobodan Djukanovi{\'c}, and Milo{\v{s}} Dakovi{\'c},
\newblock ``Instantaneous frequency in time--frequency analysis: Enhanced
  concepts and performance of estimation algorithms,''
\newblock {\em Digital Signal Processing}, vol. 35, pp. 1--13, 2014.

\end{thebibliography}

\end{document}